\documentclass[prb,floatfix,superscriptaddress]{revtex4}
\usepackage{amsmath}
\usepackage{graphicx}
\usepackage{epstopdf}
\usepackage{times}
\usepackage{color}

\usepackage[colorlinks,bookmarks=false,citecolor=blue,linkcolor=red,urlcolor=blue]{hyperref}

\begin{document}

\title{Destruction of valence-bond order in a $S=1/2$ sawtooth chain with a Dzyaloshinskii-Moriya term}
\author{Zhihao Hao}
\author{Yuan Wan}
\affiliation{Department of Physics and Astronomy, Johns Hopkins
University, Baltimore, Maryland 21218, USA}
\author{Ioannis Rousochatzakis}
\affiliation{Max-Planck-Institut f{\"u}r Physik komplexer Systeme, N{\"o}htnitzer Stra{\ss}e 38, D-01187 Dresden, Germany}
\author{Julia Wildeboer}
\author{A. Seidel}
\affiliation{Department of Physics and Center for Materials Innovation,
Washington University, St. Louis, Missouri 63136, USA}
\author{F. Mila}
\affiliation{Institute of Theoretical Physics, \'Ecole Polytechnique
F\'ed\'erale de Lausanne, CH-1015 Lausanne, Switzerland}
\author{O. Tchernyshyov}
\affiliation{Department of Physics and Astronomy, Johns Hopkins
University, Baltimore, Maryland 21218, USA}

\begin{abstract}
A small value of the spin gap in quantum antiferromagnets with strong frustration makes them susceptible to nominally small deviations from the ideal Heisenberg model. One of such perturbations, the anisotropic Dzyaloshinskii-Moriya interaction, is an important perturbation for the $S=1/2$ kagome antiferromagnet, one of the current candidates for a quantum-disordered ground state. We study the influence of the DM term in a related one-dimensional system, the sawtooth chain that has valence-bond order in its ground state. Through a combination of analytical and numerical methods, we show that a relatively weak DM coupling, $0.115J$, is sufficient to destroy the valence-bond order, close the spin gap, and turn the system into a Luttinger liquid with algebraic spin correlations. A similar mechanism may be at work in the kagome antiferromagnet.
\end{abstract}
\maketitle

\section{Introduction}

\begin{figure}
  \includegraphics[width=0.5\columnwidth]{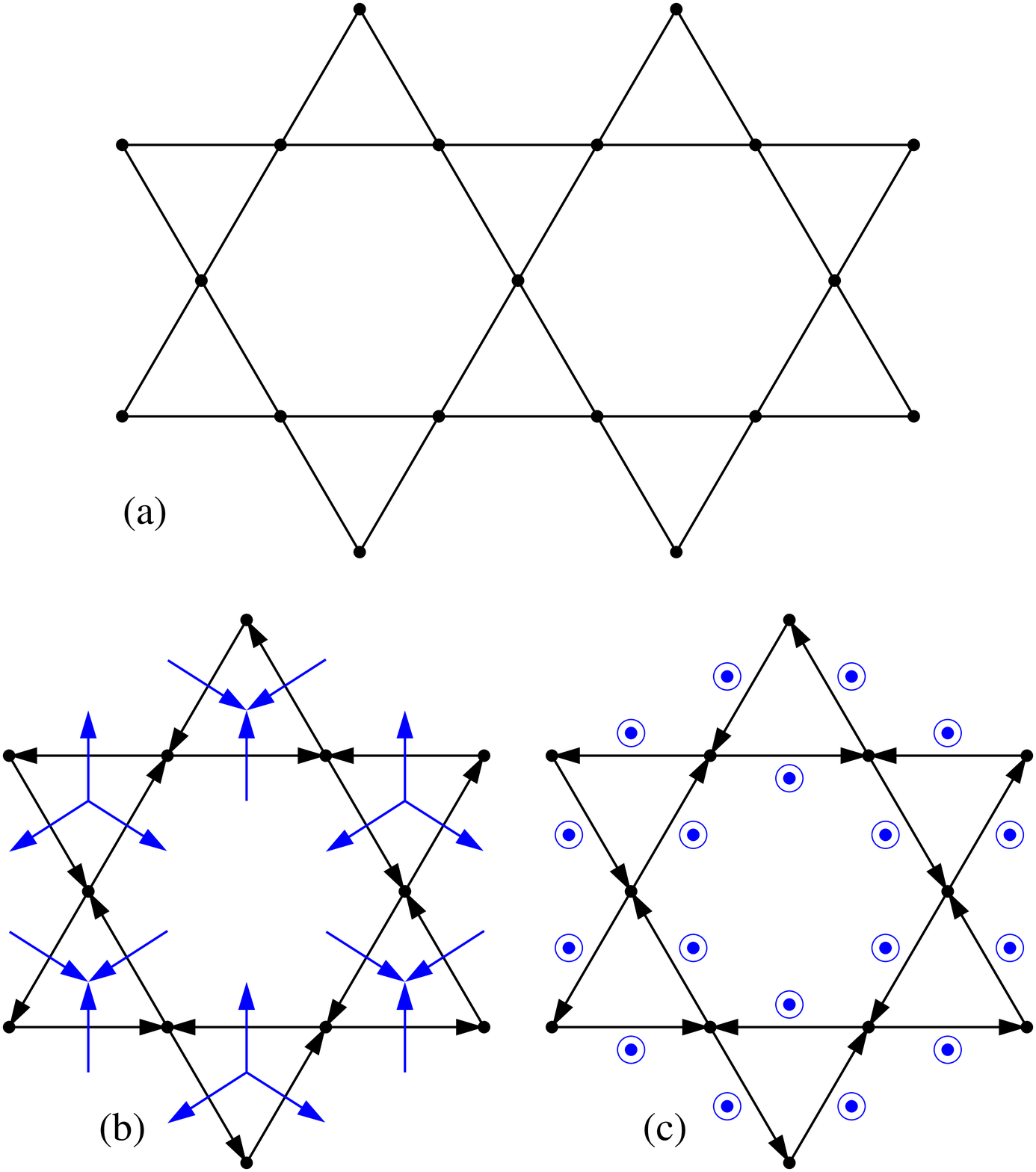}
  \caption{(a) Kagome lattice. (b) and (c) In-plane and out-of-plane components of the DM vector $\mathbf D_{ij}$ shown for directed links $(i \to j)$ on kagome. }
  \label{fig:dm}
\end{figure}

Antiferromagnets with $S=1/2$ and on non-bipartite lattices are considered viable candidates for exotic ground states and excitations. Geometrical frustration and strong quantum fluctuations tend to suppress long-range magnetic order. The resulting ground state does not break the symmetry of global spin rotations, but its exact properties remain subject of vigorous debate, with proposals ranging from valence-bond crystals that break some lattice symmetries\cite{marston:5962,nikolic:024401,PhysRevB.63.014413} to valence-bond liquids that fully preserve the symmetry of the Hamiltonian.\cite{ryu:184406,ran:117205,hermele:224413, PhysRevB.80.165131} A spin-liquid state with an energy gap to all excitations may further possess a hidden topological order. Several antiferromagnetic materials
without long-range magnetic order well below the characteristic Curie-Weiss temperature scale have been discovered recently, most notably herbertsmithite Cu$_3$Zn(OH)$_6$Cl$_2$,\cite{RSW} where no magnetic order has been detected down to 50 mK,\cite{helton:107204,vries:157205,imai:077203,ofer-2006,olariu:087202} even though the exchange interaction is estimated to be $J = 180$ K.  The material is a ``structurally perfect''\cite{JACS.127.13462,helton:107204} realization of the $S=1/2$ Heisenberg antiferromagnet on kagome, a network of corner-sharing triangles, Fig.~\ref{fig:dm}(a).

While most of the theoretical studies of quantum antiferromagnets deal with the pure Heisenberg model with nearest-neighbor exchange, real systems inevitably deviate from this idealization. Frustrated magnets in particular are sensitive to various nominally weak perturbations.  In this paper, we deal with the Dzyaloshinskii-Moryia (DM) interaction,\cite{Dzyaloshinsky, PhysRev.120.91} the antisymmetric version of the Heisenberg exchange induced by the spin-orbit coupling.  The Hamiltonian of such a system is
\begin{equation}
H = \sum_{\langle ij \rangle}
    [J \, \mathbf S_i \cdot \mathbf S_{j}
    + \mathbf{D}_{ij}\cdot (\mathbf{S}_i \times \mathbf{S}_j)].
    \label{eq:H}
\end{equation}
In herbertsmithite, the DM term is allowed by the crystal symmetry.  The in-plane and out-of-plane components of the DM vector $\mathbf D_{ij}$ on kagome are shown in Fig.~\ref{fig:dm}(b) and (c).   From eSR measurements,\cite{zorko:026405} the DM vector has the magnitude $D = 0.08J$ and is dominated by the out-of-plane component, whereas the in-plane component is small, $D_\mathrm{in} = 0.01J \pm 0.02J$. The DM term can be gauged away by an appropriate rotation of the local spin axes,\cite{Perk1976115,PhysRevLett.69.836} provided that its ``line integral'' vanishes for any closed loop $abc \ldots yza$:
\begin{equation}
\mathbf D_{ab} + \mathbf D_{bc} + \ldots + \mathbf D_{yz} + \mathbf D_{za} = 0.
\label{eq:loopD}
\end{equation}
It can be seen from Fig.~\ref{fig:dm}(b) that the in-plane component satisfies Eq.~(\ref{eq:loopD}) and thus can be gauged away. The out-of-plane component cannot be removed in this way and thus represents a physical perturbation. In this work, we concentrate on the out-of-plane component of $\mathbf D$.

A growing evidence from numerical studies\cite{EuroPhysJB.2.501, singh:180407, PhysRevLett.104.187203, jiang:117203, Science.332.1173} indicates that the pure Heisenberg model, $D=0$, has a $S=0$ ground state with a small but finite energy gap for $S=1$ excitations, with estimates ranging from $\Delta = 0.05 J$ to $0.15 J$.  These values are comparable to the strength of the DM term, so it is plausible that the low-energy properties of herbertsmithite are influenced by the DM interaction.

The effects of the DM interaction on the kagome antiferromagnet were first studied by Rigol and Singh\cite{PhysRevLett.98.207204,PhysRevB.76.184403} in order to explain low-temperature paramagnetism in herbertsmithite: an upturn in magnetic susceptibility at low temperatures\cite{PhysRevLett.104.147201} seems to indicate the absence of a spin gap. Tovar \textit{et al.}\cite{PhysRevB.79.024405} concluded that a finite DM term could be responsible for the non-zero susceptibility observed in experiment even if the spin gap remains finite. A study employing exact diagonalization \cite{cepas:140405} showed that a sufficiently strong DM term, $D > D_c \approx 0.10J$, induces long-range magnetic order in the ground state, with magnetic moments lying in the plane. This was later confirmed by employing the Schwinger-boson approach.\cite{PhysRevB.81.064428, PhysRevB.81.144432} The ordering tendency is easy to understand by turning to the classical variant of the Heisenberg model.  There, the
 out-of-plane $\mathrm{D}$ vectors shown in Fig.~\ref{fig:dm}(c) lift the extensive degeneracy of the classical ground states leaving a $\mathbf q=0$ ground state that spontaneously breaks the remaining O(2) symmetry of the DM Hamiltonian (\ref{eq:H}). Later numerical work\cite{PhysRevB.79.214415} turned up some evidence that the system may have an intermediate phase between $D_{c1} \approx 0.05 J$ and $D_{c2} \approx 0.10J$, where $S_z=1$ excitations become gapless but the spin O(2) symmetry remains intact.  In the absence of an obvious order parameter that would uniquely identify the intermediate phase, the authors of Ref.~\onlinecite{PhysRevB.79.214415} concluded that the appearance of an intermediate phase might be a finite-size effect.  Further work in this direction is required to elucidate the nature---and even the existence---of the intermediate phase and its possible relevance to herbertsmithite.

In our previous work,\cite{PhysRevLett.103.187203} we have shown that the $S=1/2$ Heisenberg antiferromagnet on kagome can be viewed as a collection of fermionic spinons---topological defects with $S=1/2$---moving in an otherwise inert vacuum of valence bonds.  The spinons interact with an emerging compact U(1) gauge field whose quantized electric flux is related to the valence-bond configuration through Elser's arrow representation.\cite{PhysRevB.48.13647}  Spinons carry one unit of the U(1) charge against a negatively charged background.  These features are reminiscent of the picture of fermionic spinons proposed earlier by Marston \textit{et al.}\cite{marston:5962, ma:027204} and Hastings\cite{PhysRevB.63.014413}, who used the Abrikosov-fermion representation for spin operators.  It is worth pointing out that the Fermi statistics of spinons is not postulated \textit{ad hoc} in our approach but rather arises naturally as the Berry phase of valence bonds that are moved in th
 e process of spinon exchange. We have further shown that strong, exchange-mediated attraction binds spinons into small and heavy $S=0$ pairs and that low-energy $S=1$ excitations result from breaking up a pair into ``free'' spinons.  Thus the spin gap is determined mostly by the binding energy of a pair, which we estimated to be $0.06 J$.

From this perspective, one potential route to the closing of the spin gap could be via the destruction of the two-spinon bound state in the presence of a sufficiently strong DM term. That, however, appears unlikely for two reasons. First, the factors setting the pair binding energy---the spinon hopping amplitude and the strength of exchange-mediated attraction--are both of order $J$, so it is hard to see how a fairly weak coupling $D = 0.05 J$ to $0.10 J$ can disrupt the pairing. Second, a quantum phase transition to a state with long-range magnetic order can be viewed as Bose condensation of magnons,\cite{bec-review:2008} quasiparticles with $S_z=1$ and there are no low-energy excitations of this kind in the pure Heisenberg model.  Although one could think of condensing pairs of spinons with $S_z=1$, this route runs into another difficulty: such an object would carry a double U(1) charge, whereas a magnon is expected to be neutral.  Put simply, a pair of spinons is a topological defect whose motion affects the valence-bond background, which is uncharacteristic of magnon motion.

A possible way out is to postulate that the condensing objects are pairs consisting of a spinon and its antiparticle. Such a composite object would have zero U(1) charge and be topologically trivial, like a magnon. In the pure Heisenberg model, the energy cost of creating a spinon and its antiparticle is approximately $0.25J$.\cite{PhysRevLett.104.177203} As we will see, the DM term lowers the kinetic energy of both spinons and their antiparticles. It is thus reasonable to expect that, at some critical coupling strength $D_c$, the energy cost of adding a pair vanishes.

\begin{figure}
\includegraphics[width=0.5\columnwidth]{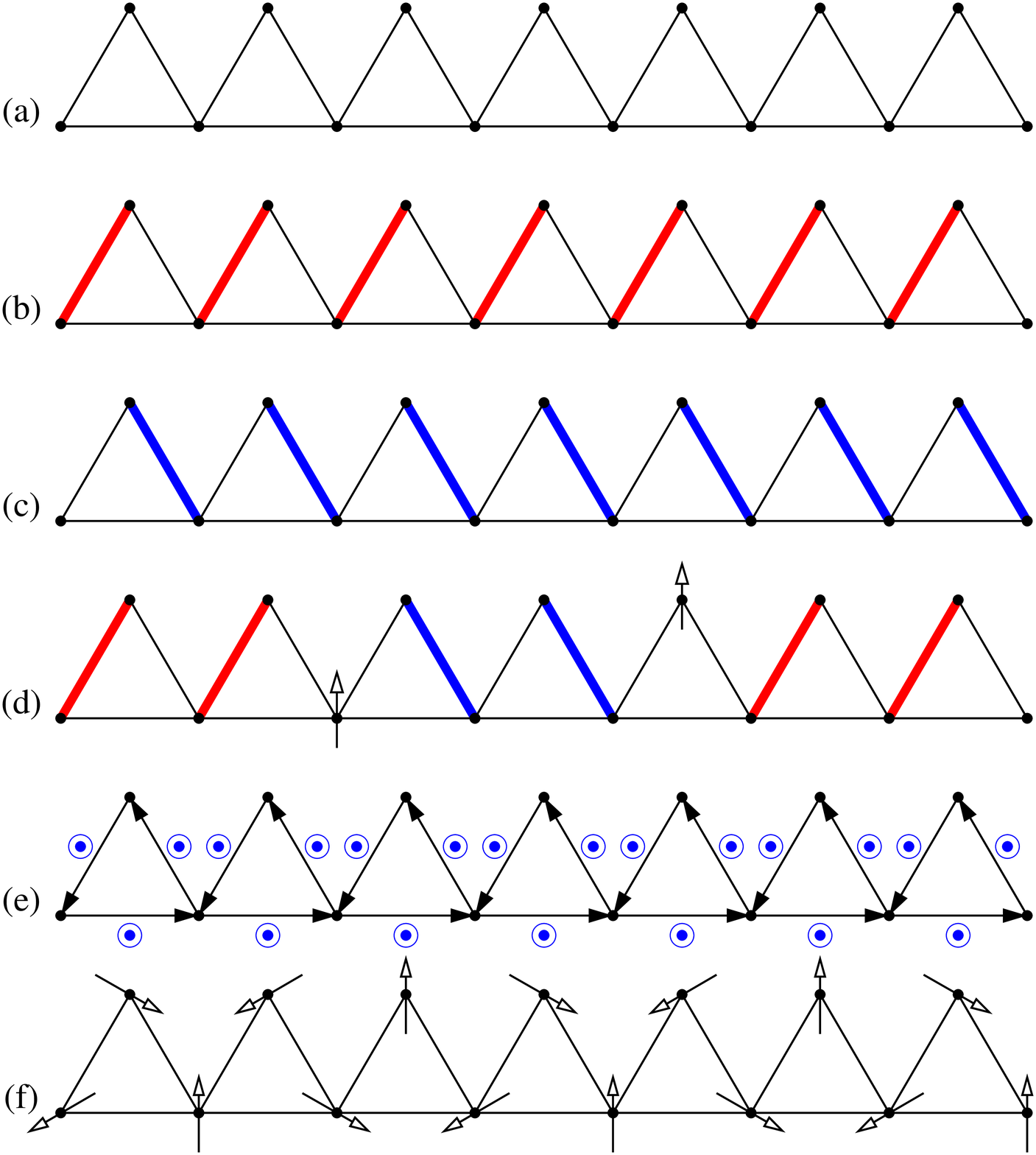}
\caption{(a) The sawtooth chain. (b) and (c) Its valence-bond ground states. (d) Spin-1/2 excitations: kink (left) and antikink (right). (e) Orientation of the DM vectors $\mathbf D_{ij}$. (f) The ground state of the classical model has a commensurate magnetic order with the wavenumber $q/2\pi = - 1/3$.}
\label{fig:sawtooth}
\end{figure}

To test this scenario, we have studied a toy version of the kagome antiferromagnet known as the sawtooth spin chain,\cite{PhysRevB.53.6393, PhysRevB.53.6401} a one-dimensional lattice of corner-sharing triangles, Fig.~\ref{fig:sawtooth}(a). To make a connection with kagome, exchange couplings are set equal for all bonds. At $D=0$, the chain has two valence-bond ground states, Fig.~\ref{fig:sawtooth}(b) and (c), that spontaneously break the mirror reflection symmetry. Spin excitations are topological defects: domain walls with spin $S=1/2$, Fig.~\ref{fig:sawtooth}(d). The domain walls come in two flavors: kinks have zero energy and are localized, whereas antikinks are mobile and have a minimum energy of $0.215J$.\cite{PhysRevB.53.6393} These excitations can only be created in pairs by a local perturbation acting in the bulk.  As we discussed elsewhere,\cite{PhysRevB.81.214445} spinons of the kagome antiferromagnet have similar properties, with one notable exception: the ground state of the sawtooth chain is free from the defects, whereas kagome has a finit
 e concentration of antikinks (1/3 per site) bound into $S=0$ pairs.

We have studied the sawtooth spin chain with exchange and Dzyaloshinskii-Moriya interactions, Eq.~(\ref{eq:H}). The $\mathbf D_{ij}$ vectors had the same length and a uniform out-of-plane orientation preserving the translational symmetry of the chain as shown in Fig.~\ref{fig:sawtooth}(e).  Qualitatively similar results were obtained for the staggered choice of $\mathbf D_{ij}$, but we will not provide the details here. The introduction of the DM term preserves the mirror symmetry of the Hamiltonian (it inverts the $x$ coordinate of the lattice and the $S_y$ and $S_z$ components of the spins), so that the notion of a valence-bond order that spontaneously breaks this symmetry is still valid.  The valence-bond order survives to a finite value of the DM coupling.

As described below, kinks become mobile in the presence of a DM term.  Their minimal energy becomes negative, growing linearly with $D$. The minimal energy of an antikink remains unchanged to the first order in $D$, so one can expect that the minimum energy of a kink-antikink pair will vanish when $D$ reaches a critical value $D_c$ of the order of the initial spin gap, $0.215 J$.  In Sec.~\ref{sec:DOQ}, we describe a calculation of the spinon spectrum in the presence of a nonzero $D$, from which we obtained an estimate of the critical DM strength, $D_c = 0.087 J$.  For $D>D_c$, spontaneous creation of kink-antikink pairs leads to a finite concentration of topological defects, which obliterates the valence-bond order and restores the reflection symmetry of the lattice.  This scenario is reminiscent of quantum phase transition at the end of magnetization plateaus in the $S=1/2$ Ising-Heisenberg chain\cite{PhysRevB.18.421} and in a frustrated two-leg ladder.\cite{fouet:2006prb}. In both of those models, the condensation of domain walls turns a state with a broken translational symmetry and gapped excitations into a gapless phase with incommensurate spin correlations decaying as a power of the distance.  Exact diagonalization calculation for the sawtooth chain with DM interactions, described in Sec.~\ref{sec:ED}, are consistent with this scenario.

\section{Spinon dispersions}\label{sec:DOQ}

\subsection{$D=0$}

We briefly review the physics of the sawtooth chain in the pure Heisenberg model without the DM term.\cite{PhysRevB.53.6393, PhysRevB.53.6401, PhysRevB.81.214445} The Hamiltonian of the system is
\begin{equation}\label{exchange}
H = J \sum_{\langle ij\rangle} \mathbf S_i \cdot \mathbf S_j
	= \frac{J}{2}\sum_{\Delta}\left(\mathbf{S}_{\Delta}^2-9/4\right),
\end{equation}
where the $\mathbf{S}_{\Delta}$ is the total spin of triangle $\Delta$. The energy is minimized when $S_\Delta = 1/2$ for every triangle, which can be achieved by putting a singlet bond on every triangle. The ground state is doubly degenerate. The two ground states shown in Fig.~\ref{fig:sawtooth}(b) and (c) violate the symmetry of reflection.

Two types of domain walls interpolate between the ground states: the kink and the antikink, Fig.\ref{fig:sawtooth}(d). A kink is an excitation with zero energy that happens to be an exact eigenstate of the Hamiltonian (\ref{exchange}). Thus kinks are localized in the exchange-only model. The localized nature of kinks can be traced to an accidental degeneracy of the ground state of the exchange Hamiltonian on a triangle with half-integer spins in addition to the two-fold Kramers degeneracy. The two degenerate states with Sz=1/2 have spin current going clockwise or counter clockwise around the triangle. The states also carry electric currents of opposite directions.\cite{PhysRevB.78.024402} An alternative set of basis states would have distinct valence-bond averages $\langle \mathbf{S}_i\cdot \mathbf{S}_j\rangle$ on the three bonds, which translates to nonzero electric charge on the three sites.\cite{PhysRevB.78.024402}

In contrast, an antikink is mobile.  The motion of an antikink is accompanied by the emission and absorption of kink-antikink pairs.  The existence of a finite spin gap guarantees that these excitations are virtual.  Polarization effects can be taken into account by using a variational approach.  At the crudest level, the Hamiltonian (\ref{exchange}) is projected onto the Hilbert space with a single antikink to obtain an effective hopping Hamiltonian for an antikink:
\begin{equation}\label{eq:n=0}
    H^{(1)}|x\rangle
    = \frac{5J}{4} |x\rangle
    - \frac{J}{2} |x+1\rangle
    - \frac{J}{2} |x-1\rangle.
\end{equation}
where $|x\rangle$ is a state with an antikink on triangle $x$. The energy dispersion of the antikink is
\begin{equation}
E_a(k) = 5J/4 - J\cos{k},
\end{equation}
with the minimum energy $\Delta = 0.25J$.  In view of the zero energy of a kink, this value is the spin gap.

This estimate can be further improved by enlarging the Hilbert space to include virtual excitations in the immediate neighborhood of an antikink. This yields an improved estimate of the spin gap, $\Delta = 0.219 J$,\cite{PhysRevB.81.214445} which is quite close to the result obtained by exact diagonalization, $\Delta = 0.215J$.\cite{PhysRevB.53.6393}

It seems clear from the above that the variational approach provides a reliable description of the low-energy spin excitations in the pure Heisenberg model. We will use the lowest-order approximation for $D \neq 0$, without correcting for the vacuum polarization, to obtain a rough estimate for the critical coupling $D_c$.

\subsection{$D\neq0$}

In the presence of a nonzero DM term, kinks become mobile. For a single triangle, this means the splitting of the accidental degeneracy mentioned previously: the energy of a state with $S_z=+1/2$ now depends on the orbital momentum, reflecting the spin-orbit origin of the DM term.

For an infinite chain, we follow the variational method described above and work in the Hilbert space spanned by states $|x\rangle$ with a single kink located between triangles $x$ and $x+1$. These states are not orthogonal to each other because they are not eigenstates of the same Hermitian operator. The overlap is
\begin{equation}
\langle x_1|x_2\rangle = 2^{-|x_{1} - x_{2}|}.
\end{equation}
As with antikinks,\cite{PhysRevB.81.214445} a simple rotation can be made to obtain an orthonormal basis $\{|\tilde{x}\rangle\}$:
\begin{equation}\label{orthonormal}
    |\tilde{x}\rangle=\frac{2}{\sqrt{3}}|x\rangle - \frac{2}{\sqrt{3}}|x-1\rangle.
\end{equation}
The matrix elements of the effective Hamiltonian in this subspace are
\begin{equation}\label{hamikink}
    \langle \tilde{x}_1|H|\tilde{x}_2\rangle
     = -\frac{3iD}{2} \, 2^{-|x_{1}-x_{2}|} \, \mathrm{sgn}(x_1-x_2),
\end{equation}
where the sign function is defined in such a way that $\mathrm{sgn}(0) = 0$.  A Fourier transform of the matrix element yields the energy dispersion of the kink:
\begin{equation}\label{dispkink}
    E_\mathrm{k}(k)= \frac{6D\sin{k}}{5-4\cos{k}}.
\end{equation}
The bottom of the band is at $E_\mathrm{k}^\mathrm{min} = -2|D|$. For $D>0$, it is reached for an incommensurate wavenumber $k/2\pi = -\mathrm{acos}{(4/5)}/2\pi \approx -0.10$.

The calculation of the antikink case proceeds in a similar way.  The basis states $\{|x\rangle\}$, with an antikink located at triangle $x$, can be orthogonalized in the same way to yield an orthonormal basis $\{|\tilde{x}\rangle\}$. The matrix element of the DM term is
\begin{equation}\label{hamiantikink}
    \langle \tilde{x}_1|H_{DM}|\tilde{x}_2\rangle =
    -i D\,  2^{-|x_{1}-x_{2}|} \, \mathrm{sgn}(x_1-x_2)
    \left[\frac{3}{2}-\frac{2}{3}(\delta_{x_{1},x_{2}+1}+\delta_{x_{1},x_{2}-1})\right].
 \end{equation}
The resulting antikink dispersion is
\begin{equation}
    E_\mathrm{a}(k) = 5J/4 - J\cos{k}
    + \frac{5D}{6}\sin{k}
    +\frac{3D(4\cos{k}-1)\sin{k}}{10-8\cos{k}}.
    \label{dispantikink}
\end{equation}
For $D\ll J$, the lowest energy of an antkink
$E_\mathrm{a}^\mathrm{min} = J/4 - 14D^2/J + \mathcal O(D^4/J^3)$.
The bottom of the band is located at $k/2\pi = -8D/3\pi J + \mathcal O(D^3/J^2)$.

The above energy dispersions were computed for spinons with $S_z=+1/2$. The dispersions for $S_z=-1/2$ can be obtained by changing $k \mapsto -k$.

The bottom edge of the two-particle continuum as a function of total momentum is shown as solid lines in Fig.~\ref{fig:DispSz0} for $S_z=0$ and in Fig.~\ref{fig:DispSz1} for $S_z=+1$.  (The former is a combination of two continua, one for a kink with $S_z=+1/2$ and an antikink with $S_z=-1/2$, the other for a kink with $S_z=-1/2$ and an antikink with $S_z=+1/2$.)  The edge dispersion mostly tracks the dispersion of the heavier particle, in this case the kink (\ref{dispkink}).  The minimum energy of a kink-antikink pair
\begin{equation}\label{min}
    E^\mathrm{min} = J/4 - 2|D| - 14D^2/J + \mathcal O(D^4/J^3)
\end{equation}
vanishes when the DM coupling reaches the critical strength $D_c = 0.09J$. The total momentum of a $S_z=+1$ spinon pair with the lowest energy is $k/2\pi \approx -0.15$.  The gapless state arising at this critical point is expected to have transverse spin fluctuations with this wavenumber.  The wavenumber of longitudinal spin fluctuations is determined by the bottom of the two-spinon continuum with $S_z=0$, which occurs at $k/2\pi \approx \pm 0.06$.

\section{Exact diagonalization}\label{sec:ED}

To test the theory, we have performed an exact diagonalization study of the sawtooth chain with exchange and DM interactions.  We worked with finite chains containing $2L$ sites in a system with $L$ triangles with periodic boundary conditions. The length varied from $L=5$ to 15. Both uniform and staggered DM interactions were investigated, with qualitatively similar results. Here we report on the uniform case only. For the largest system sizes, we employed the Lanczos algorithm, which provides convergent results for the ground state energy and a limited number of low-lying excitations. To reduce the size of the Hilbert space, we used the symmetry of translations along the chain and the O(2) symmetry of spin rotations around the z-axis.

\begin{figure}
\includegraphics[width=0.32\columnwidth]{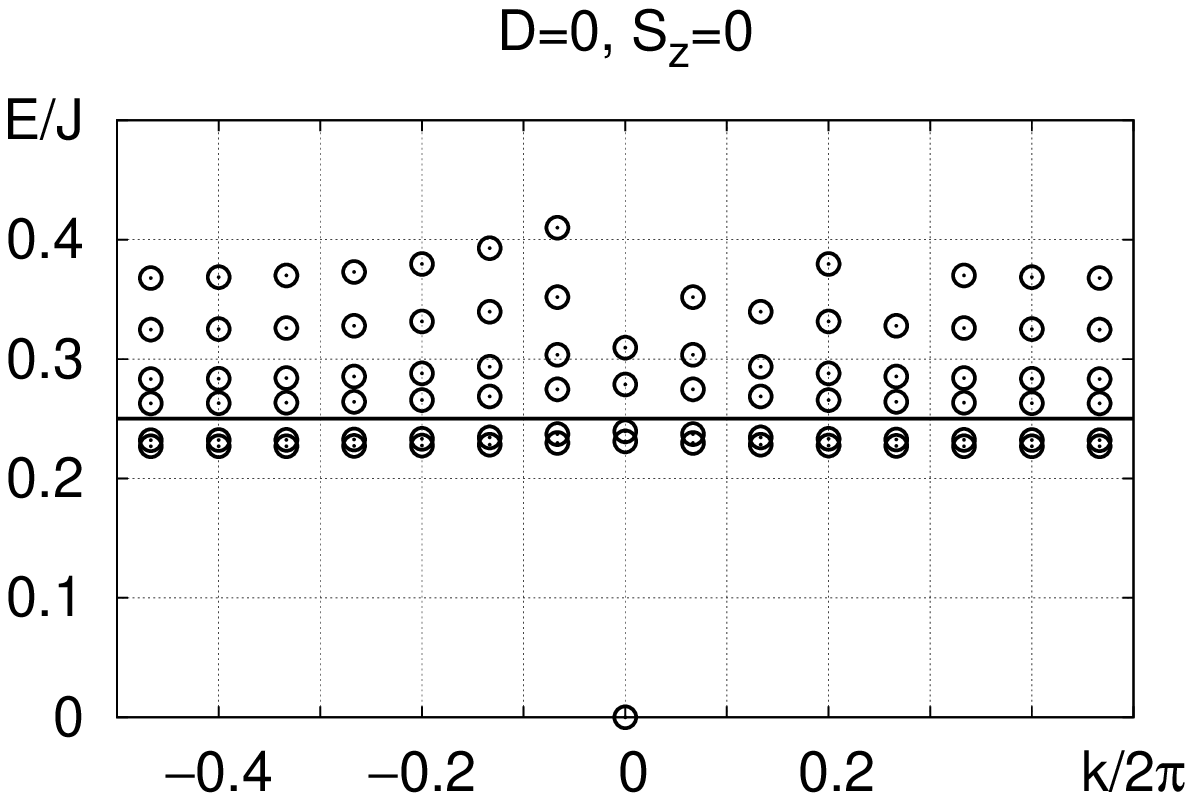}
\includegraphics[width=0.32\columnwidth]{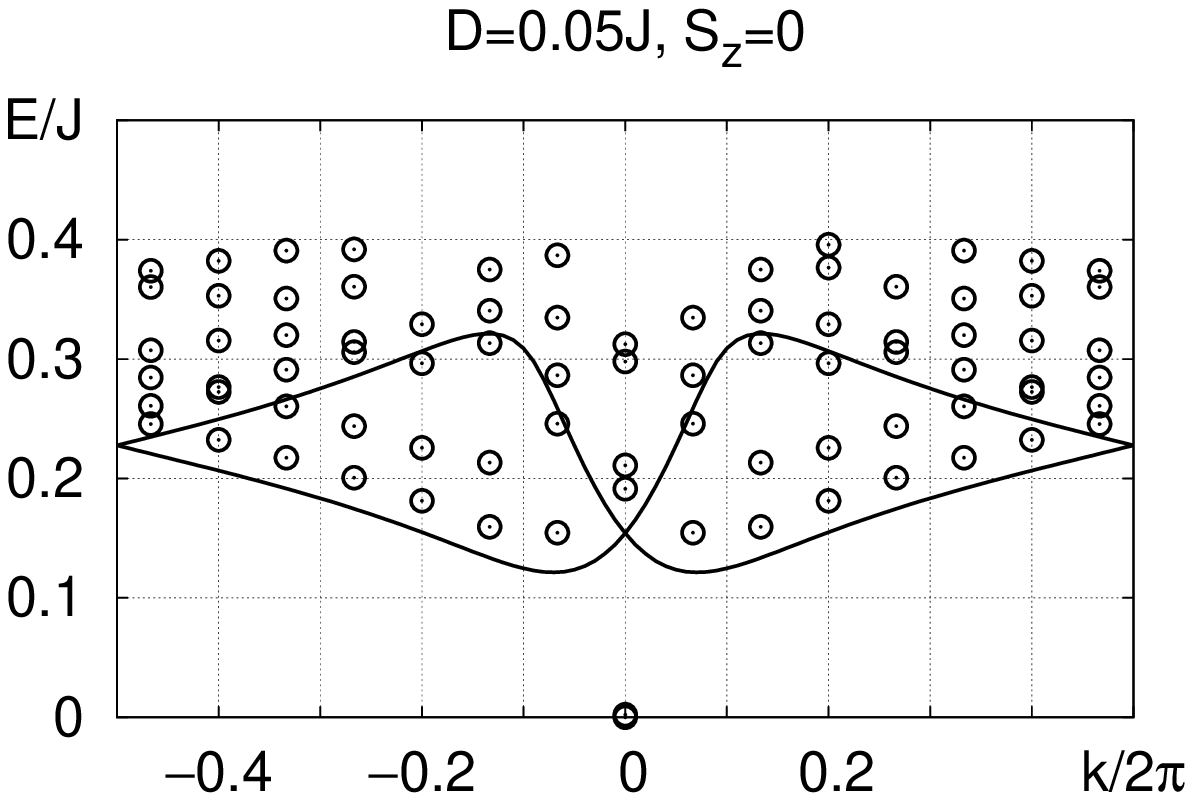}
\includegraphics[width=0.32\columnwidth]{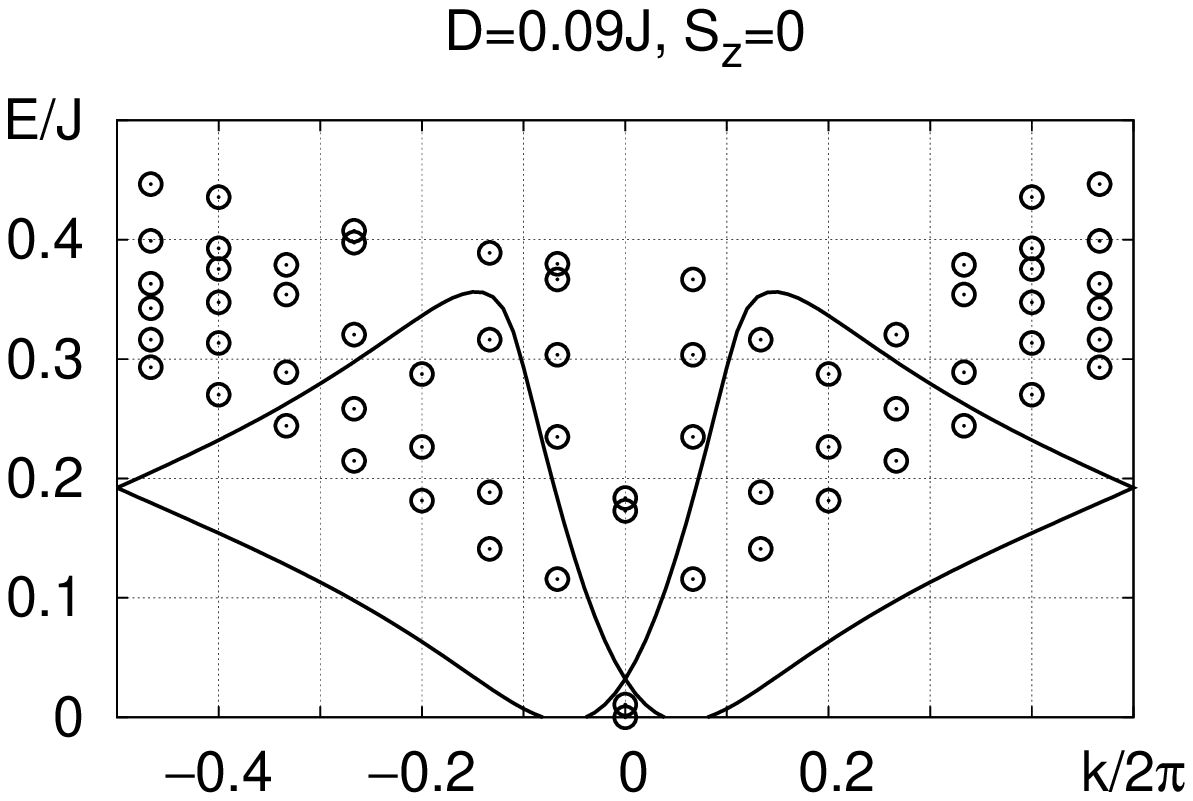}
\includegraphics[width=0.32\columnwidth]{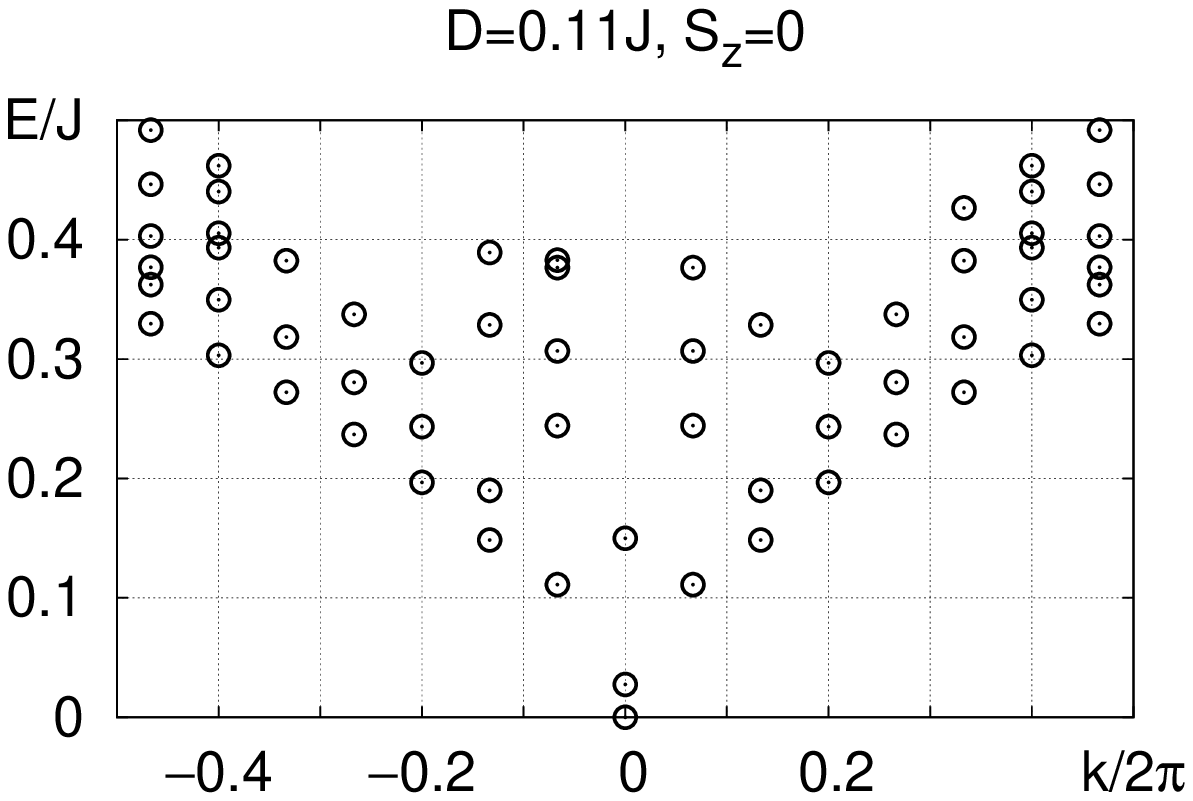}
\includegraphics[width=0.32\columnwidth]{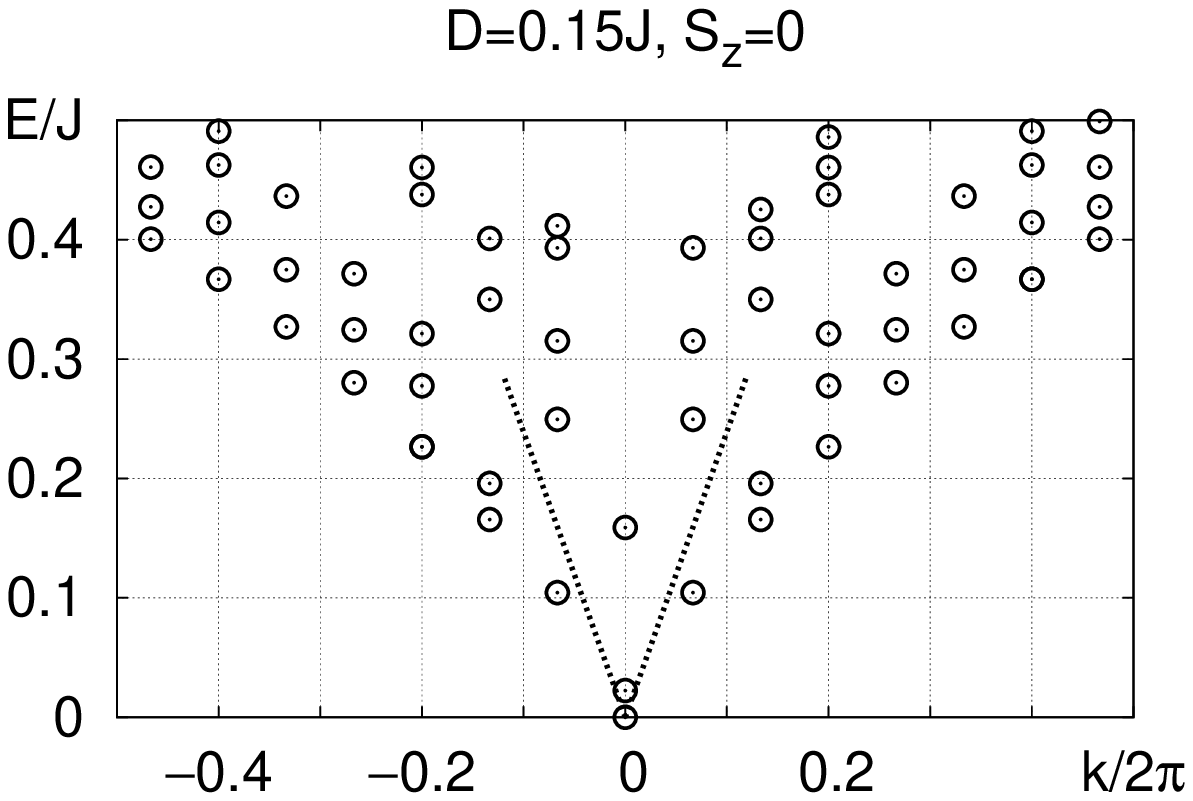}
\includegraphics[width=0.32\columnwidth]{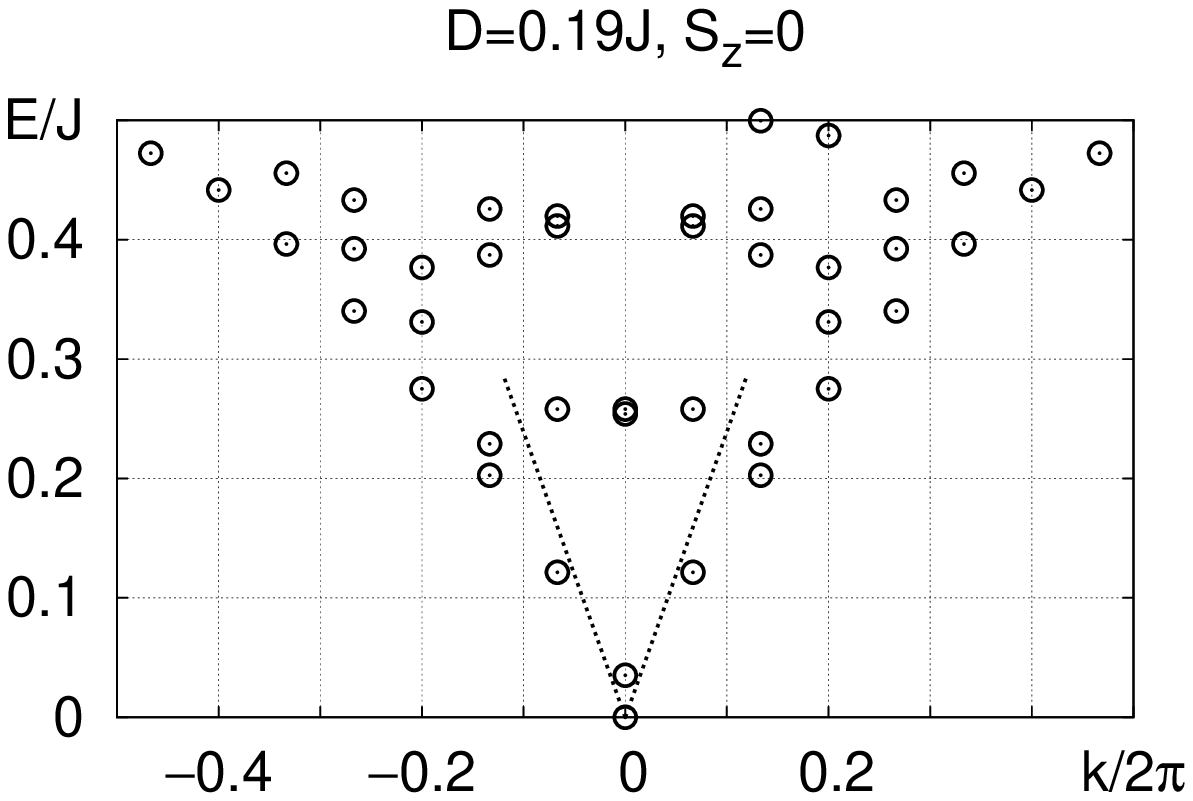}
\caption{Low-energy spectra of the sawtooth chain with a uniform DM term in the $S_z=0$ sector. Energy levels, measured relative to the ground state, are shown as a function of total momentum.  Circles are the results of exact diagonalization for a periodic chain of length $L=15$.  Solid curves show the bottoms of the two-spinon continua computed analytically.  Dashed straight lines show a linear dispersion with the speed $v = 0.36J$.}
\label{fig:DispSz0}
\end{figure}

Figure \ref{fig:DispSz0} shows the low-energy portions of the spectra in the $S_z=0$ sector for a chain with length $L=15$ (30 sites), for several values of the DM coupling $D$.  The invariance of the Hamiltonian (\ref{eq:H}) under time reversal symmetry ($S_z \mapsto -S_z$, $k \to -k$) guarantees that the $S_z=0$ spectra are symmetric under mirror reflection ($k \to -k$). The lowest-energy excitations in the $S_z=0$ sector are expected to be spinon pairs in two channels: a kink with $S_z=-1/2$ and an antikink with $S_z=+1/2$ or vice versa. The calculated edges of the two-particle continua reproduce the shape of the dispersing bottom reasonably well. However, the calculated edge shifts downward with $D$ faster than the numerical data do.

\begin{figure}
\includegraphics[width=0.32\columnwidth]{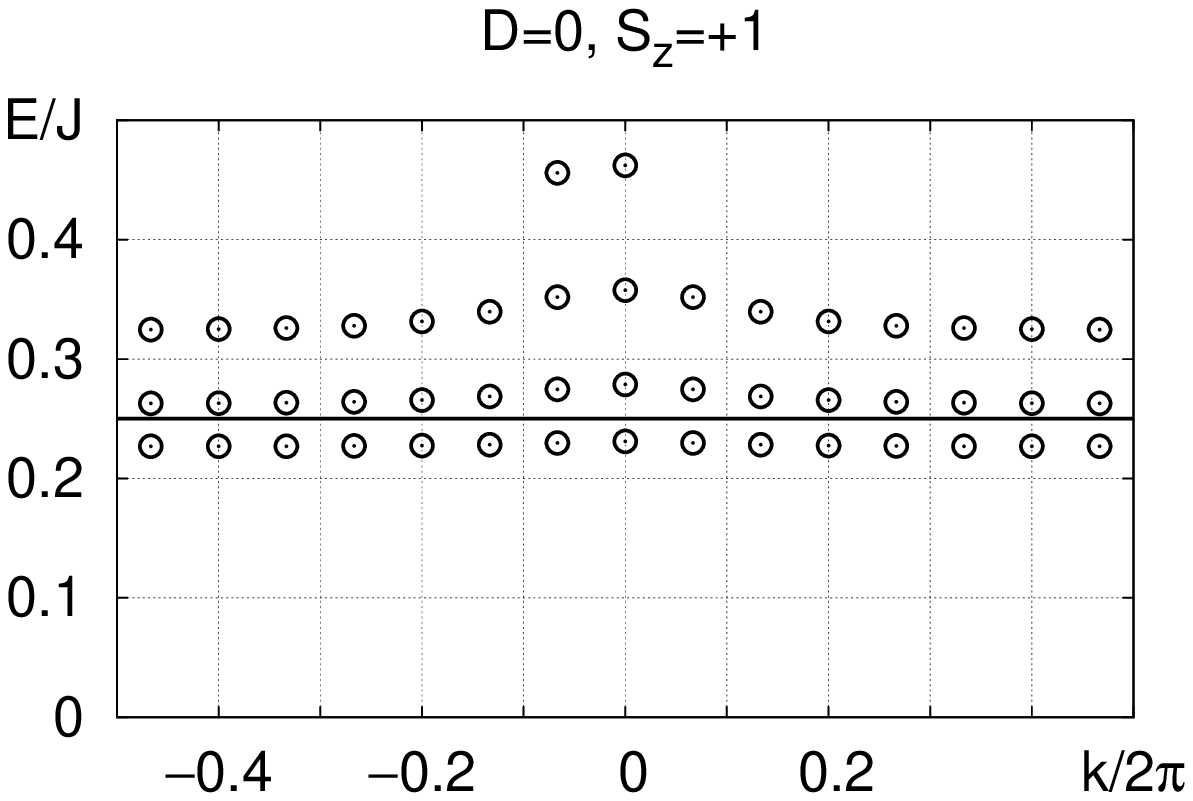}
\includegraphics[width=0.32\columnwidth]{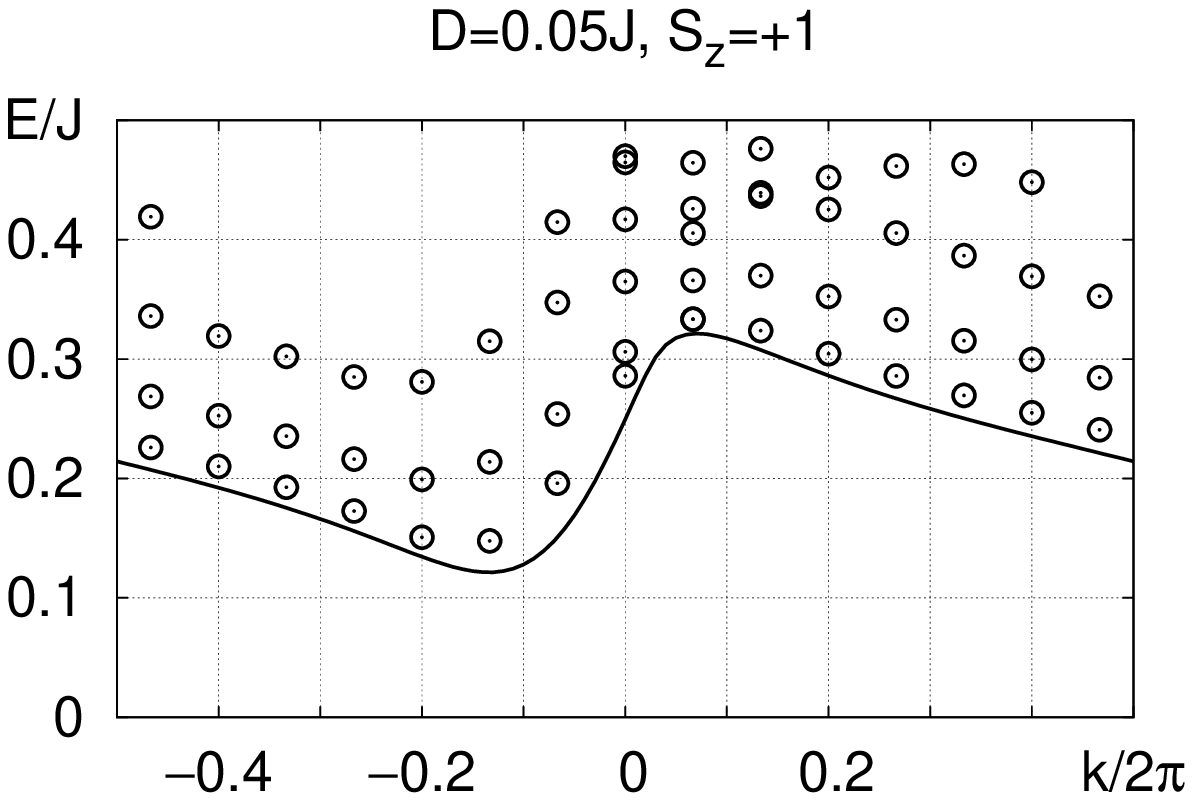}
\includegraphics[width=0.32\columnwidth]{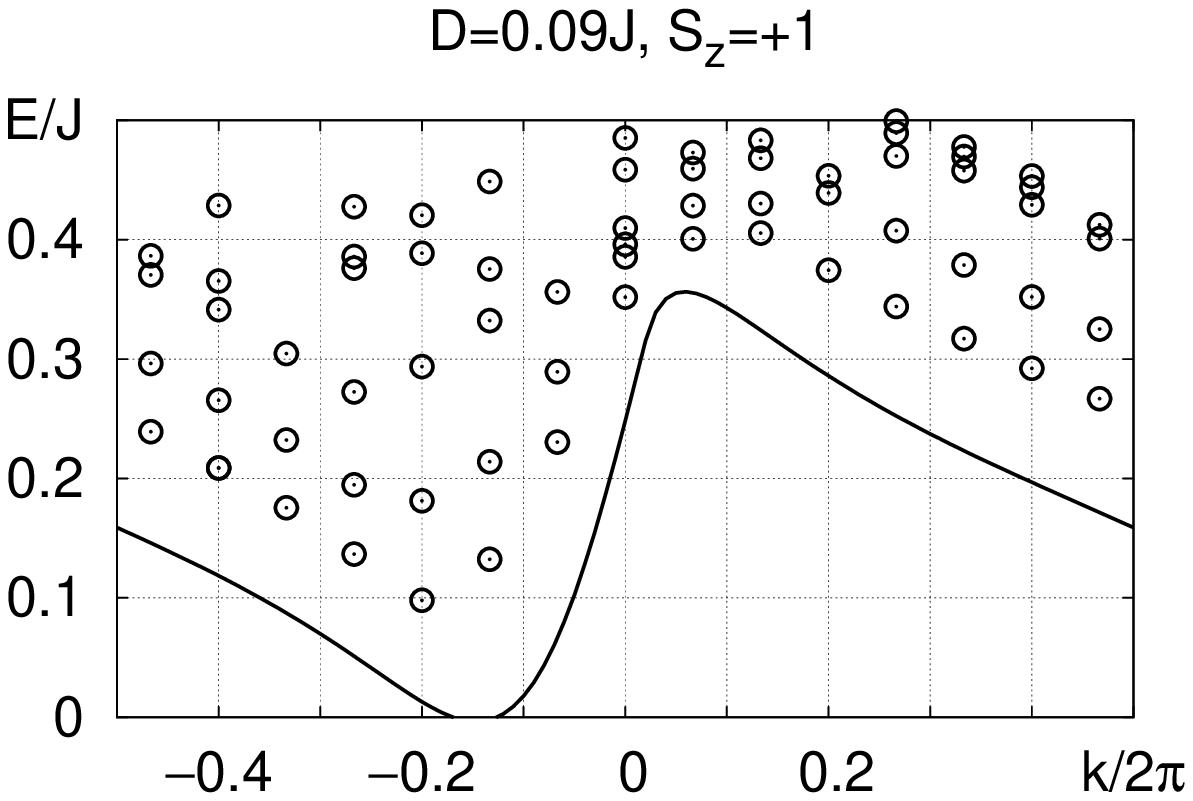}
\includegraphics[width=0.32\columnwidth]{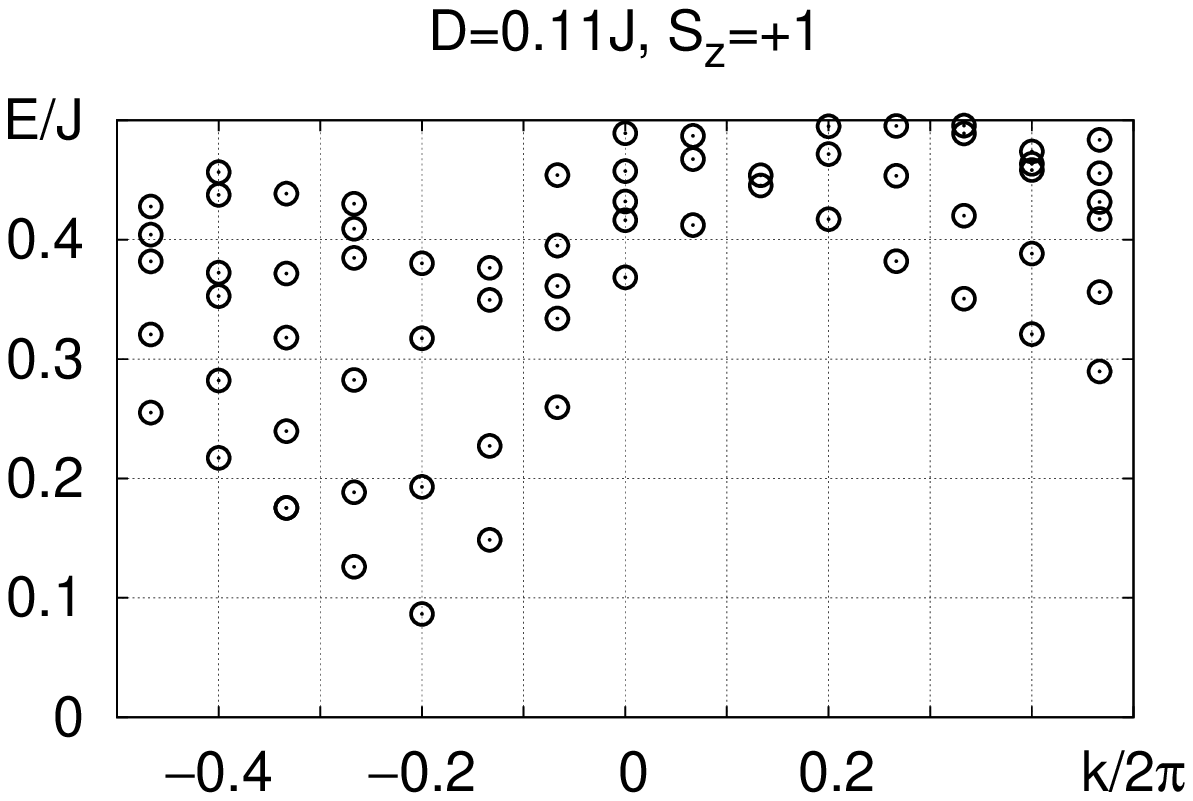}
\includegraphics[width=0.32\columnwidth]{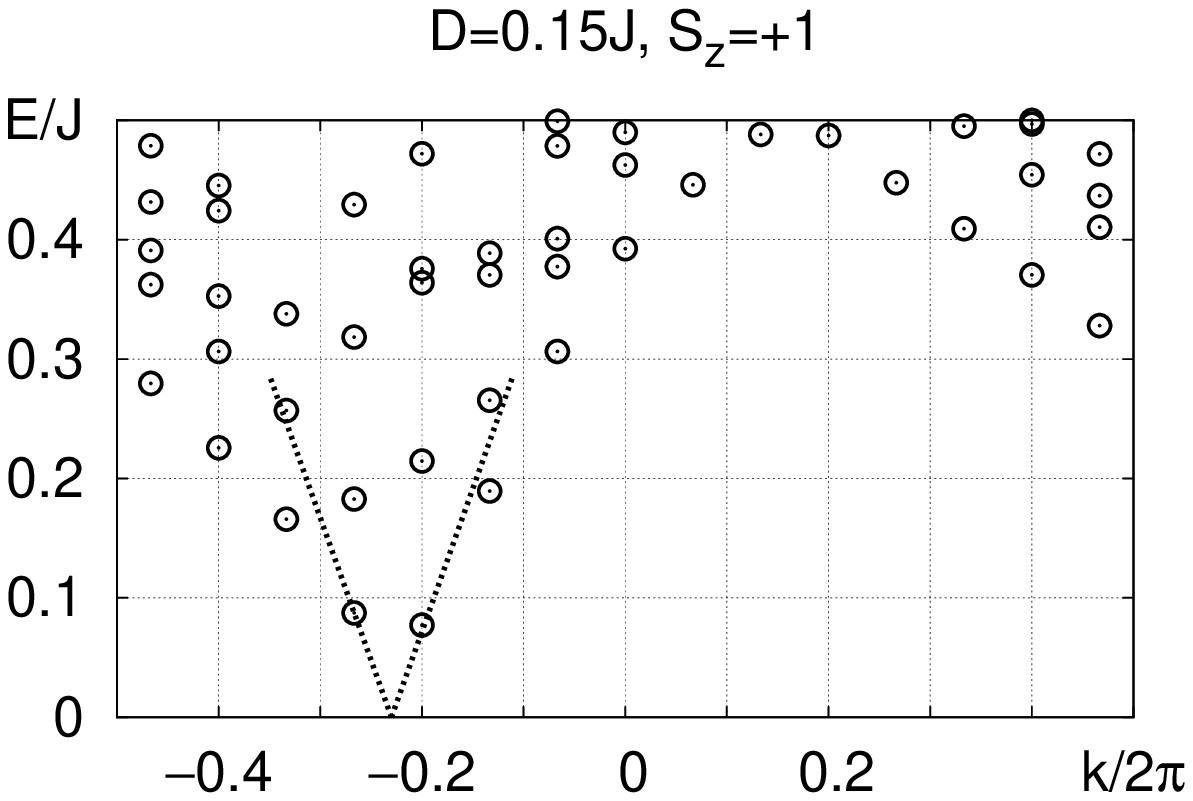}
\includegraphics[width=0.32\columnwidth]{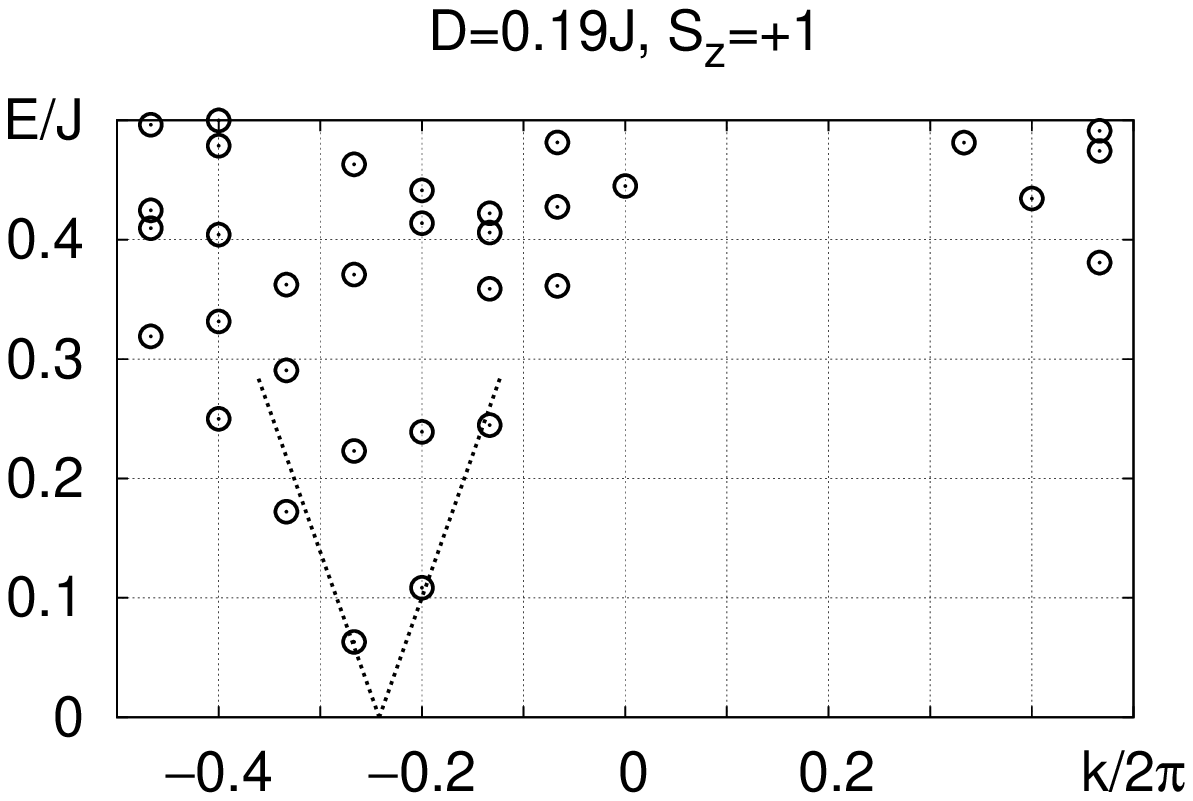}
\caption{Low-energy spectra in the $S_z=+1$ sector. Notations are the same as in Fig.~\ref{fig:DispSz0}.}
\label{fig:DispSz1}
\end{figure}

In the $S_z=+1$ sector, the spectra are not symmetric under the mirror symmetry (the $S_z=1$ spectrum maps onto that of the $S_z=-1$ sector), Fig.~\ref{fig:DispSz1}. The lowest-energy excitations are expected to be spinon pairs consisting of a kink and an antikink, both with $S_z=+1/2$. Again, the calculated bottom edge of the excitation continuum has the right shape but advances downward with $D$ somewhat too fast.  In the two-spinon approximation, both the $S_z=0$ and $S_z=1$ continua touch zero energy at $D_c = 0.09 J$. However, the numerical energy spectra appear to still have a gap at that point, see Fig.~\ref{fig:DispSz0}.

\begin{figure}
\includegraphics[width=0.32\columnwidth]{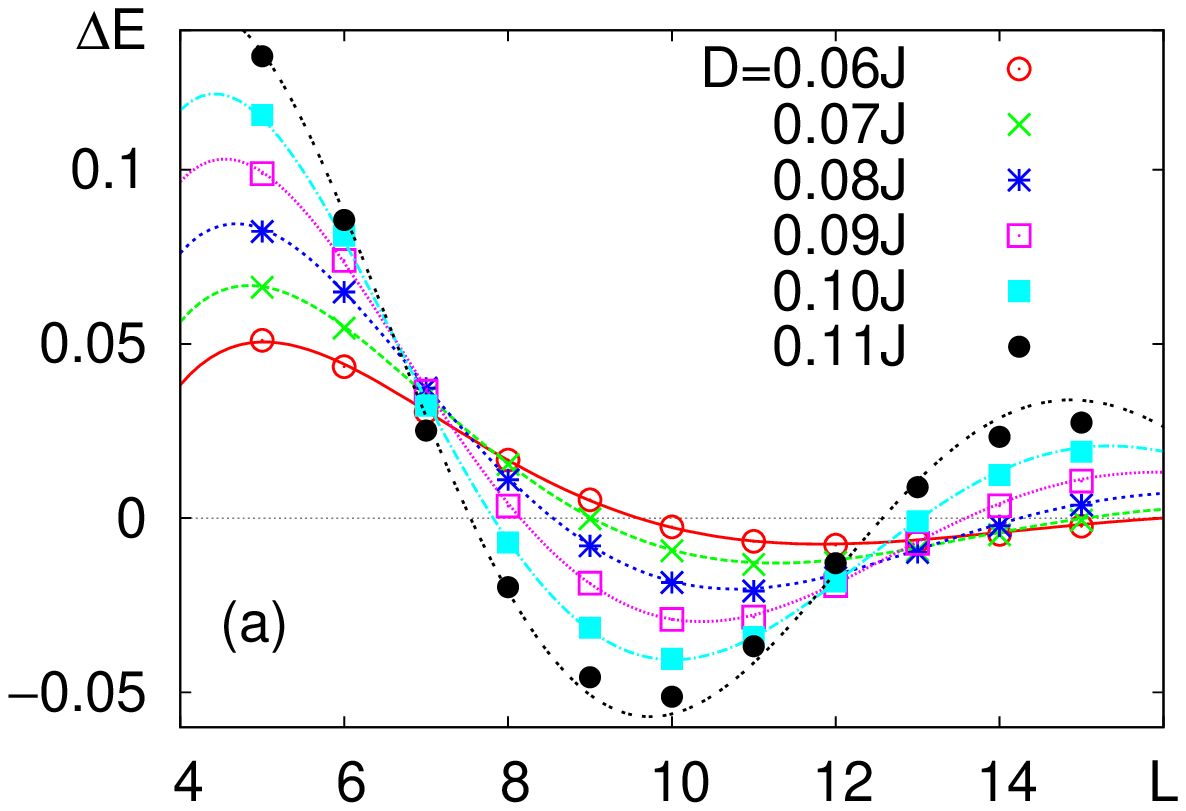}
\includegraphics[width=0.32\columnwidth]{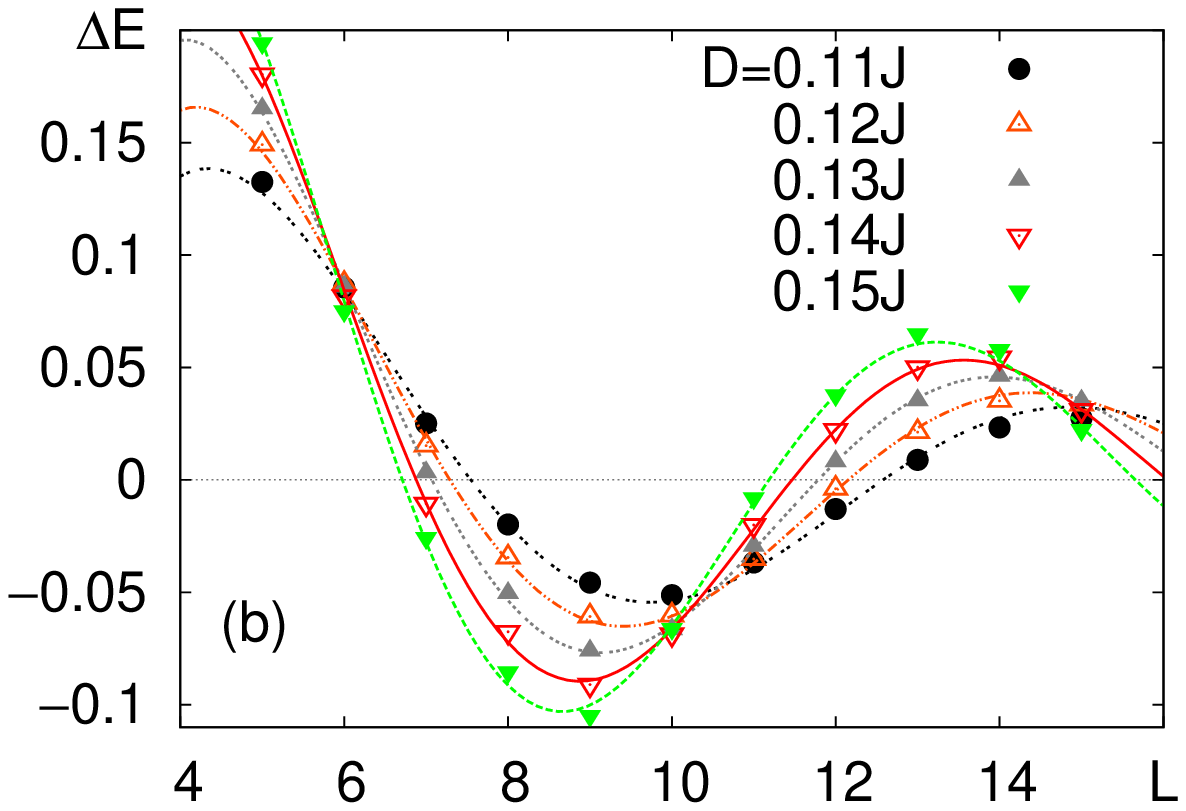}
\includegraphics[width=0.32\columnwidth]{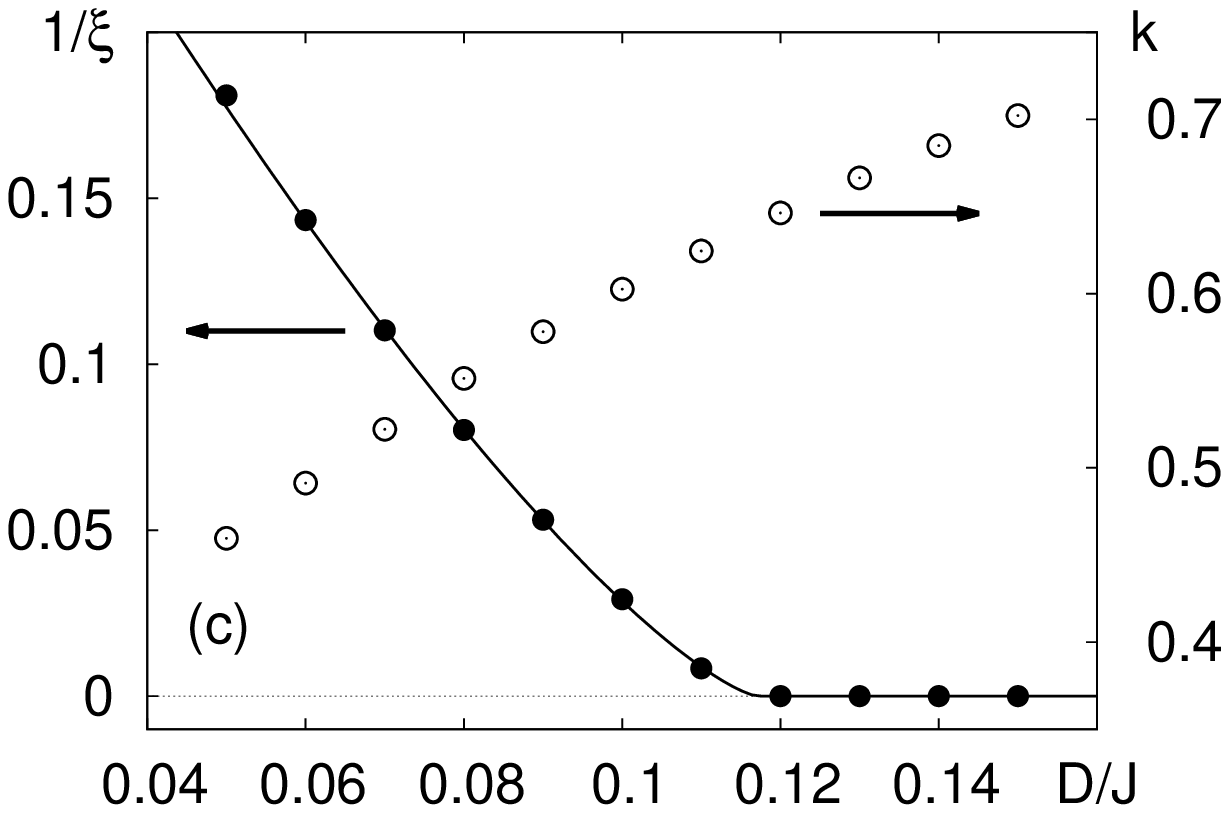}
\caption{The splitting of the ground-state doublet as a function of the system length $L$ for (a) $D<D_c = 0.115 J$ and (b) for $D>D_c$.  (c) The dependence of the inverse tunneling length $1/\xi$ and the wavenumber $k$ in the scaling form (\ref{eq:scaling}) on the DM coupling strength $D$.}
\label{fig:splitting}
\end{figure}

To locate the critical point, we turned to a scaling analysis of the ground-state splitting. In the phase with valence-bond order, the ground state is doubly degenerate in the limit $L \to \infty$. In finite systems, the ground-state doublet is split thanks to quantum tunneling.  Both members of the doublet have momentum $k=0$ because the valence-bond order preserves translational symmetry. The tunneling amplitude decays exponentially with the system length $L$ and so does the splitting.

Fig.~\ref{fig:splitting}(a) shows the splitting of the ground state for $D\leq 0.11J$. All of the data sets, with the exception of the largest coupling, are well fit by the scaling expression
\begin{equation}
\Delta E = A L^{-5/4} e^{-L/\xi} \cos{(kL)}
\label{eq:scaling}
\end{equation}
with the same prefactor $A$. The dependence of the tunneling length $\xi$ and the wavenumber $k$ is shown in Fig.~\ref{fig:splitting}(c). The tunneling length diverges, or at least greatly exceeds the maximum attainable system length $L=15$, for $D > D_c = 0.115 J$. For $0.11J \leq D \leq 0.15 J$, the finite-size dependence of the splitting was best fit by Eq.~(\ref{eq:scaling}) with $\xi=\infty$ and a $D$-dependent amplitude $A$, Fig.~\ref{fig:splitting}(b).  Apart from the oscillating factor, Eq.~(\ref{eq:scaling}) suggests a scale-invariant ground state for $D \geq D_c$. The oscillations presumably come from the interference of instantons as discussed in the Appendix.

For $D > D_c$, we expect a gapless phase with quasi-long-range incommensurate spin correlations decaying as a power of the distance.  For a sufficiently large $D$, the classical model should become a good starting point. In the classical limit, the sawtooth chain has a spiral order for any nonzero value of $D$, Fig.~\ref{fig:sawtooth}(f). Low-energy excitations are spin waves with a speed
\begin{equation}
v \approx 2.7S\sqrt{JD}.
\label{eq:v-cl}
\end{equation}
Quantum fluctuations disrupt the long-range spin order, restoring translational invariance and the O(2) symmetry. Such a phase would be a Luttinger liquid, whose lowest-energy $S_z=+1$ excitations are spin waves with a sound-like spectrum at $k_0/2\pi = -1/3$.  The numerically determined $S_z=+1$ spectra for $D \geq 0.15J$ are consistent with spin waves. At $D=0.19J$, the soft spot is located at $k_0/2\pi \approx -0.25$, not far from the classical value.  The speed of sound (estimated from the slope of the dashed lines in Fig.~\ref{fig:DispSz0} and \ref{fig:DispSz1}) is $v = 0.36J$, is not far from the classical estimate (\ref{eq:v-cl}) obtained below.

\section{Spin correlations in the ground state}

To verify the location of the quantum critical point $D_c$ and to confirm the critical nature of the ground state for $D > D_c$, we examined the long-distance behavior of spin correlations,
$G^{\alpha\beta}(r) = \langle S^\alpha(0) S^\beta(r) \rangle$,
in the ground state. In the Luttinger-liquid regime, transverse spin correlations are expected to decay as a power of the distance,\cite{giamarchi:book}
\begin{equation}
|G^{+-}(r)| \sim \frac{C}{r^{1/2K}}.
\label{eq:G-transverse}
\end{equation}
The stiffness constant $K$ varies between 1 (gas of dilute magnons) and 1/4 (gas of dilute spinons).\cite{PhysRevLett.45.1358, fouet:2006prb}

In a finite system of length $L$ with periodic boundary conditions, the Green's function depends in the same way on the chord distance \cite{JPhysC.14.2585}
\begin{equation}
d(r) = (L/\pi)\sin{(\pi r/L)}.
\label{eq:chord}
\end{equation}
In a system with $2L$ spins, this distance varies from $d \approx 1$ to $L/\pi$. In view of that, the range of distances in a system with $2 L = 30$ spins is not sufficient to reliably observe the critical behavior of the spin correlation function.

\begin{figure}
\includegraphics[width=0.33\columnwidth]{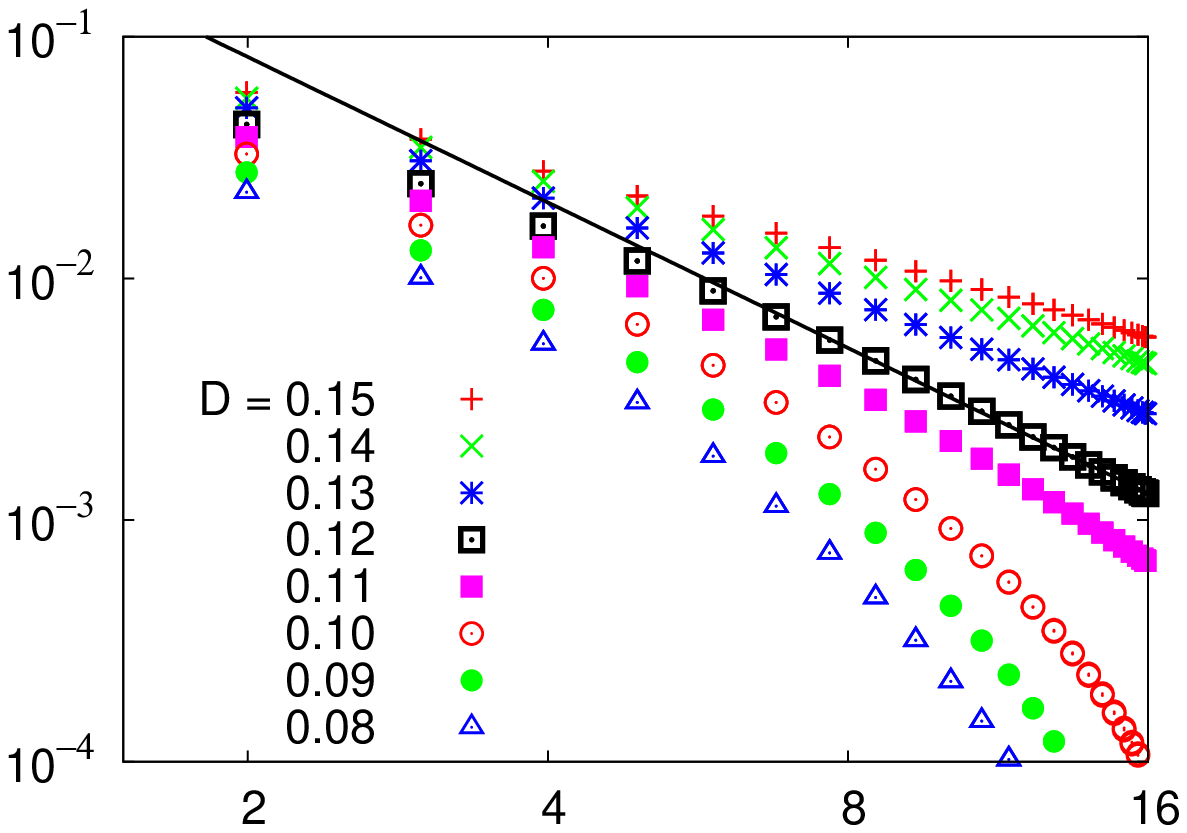}
\includegraphics[width=0.33\columnwidth]{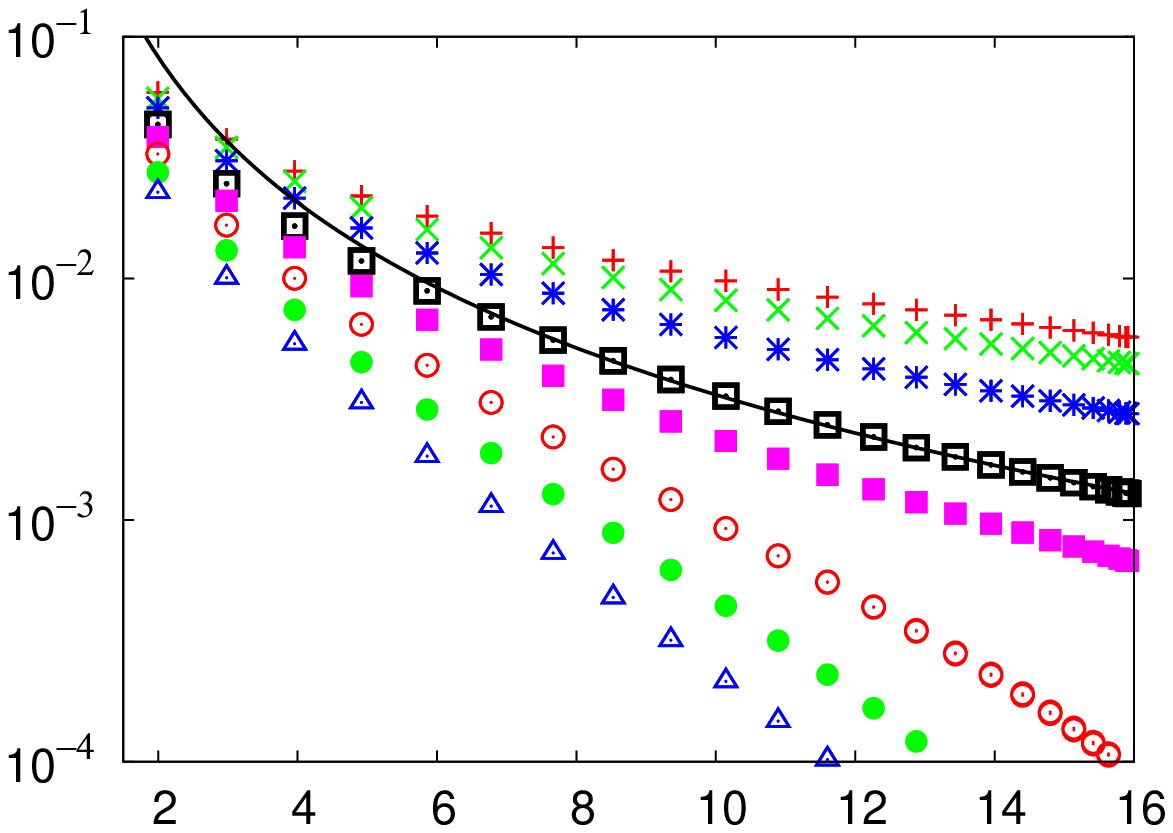}
\caption{The amplitude of transverse spin correlations (\ref{eq:G-transverse}) as a function of the chord distance (\ref{eq:chord}) on a log-log plot (left) and a simple log plot (right).}
\label{fig:DMRG}
\end{figure}

To observe the critical behavior, we used the density-matrix renormalization group (DMRG) method implemented through the Matrix Product Toolkit\cite{MPT} to obtain the ground-state wavefunction in a periodic chain with up to $2L = 100$ spins. The system has a U(1) symmetry which we took into account to reduce CPU time.
The number $m$ of states kept varied from 800 to 1200 states. Our results for the ground state energy {\it per site} for all values of DM coupling $D$ investigated are consistent with the energy per site obtained from the ED calculations.

The resulting transverse spin correlations $|G^{+-}(r)|$ in a system of length $L = 50$ are shown in Fig.~\ref{fig:DMRG} as a function of the chord distance (\ref{eq:chord}).  At largest distances $d$, the data for $D = 0.12 J$ follow a power law $C/d^2$, which is consistent with the value $K = 1/4$ at the spinon condensation point. For $D > 0.12 J$, spin correlations follow power laws with smaller slopes, indicating $K>1/4$. For $D < 0.12 J$, the power-law scaling breaks down at large $d$ changing to an exponential dependence. The estimated critical point, $D_c = 0.12 J$, is in reasonable agreement with the value $D_c = 0.115 J$ obtained from the splitting of the ground-state doublet.

\section{Discussion}

Analytical arguments and numerical evidence presented above supports the following scenario. In the absence of the Dzyaloshinskii-Moriya term, the sawtooth chain has a doubly degenerate ground state with valence-bond order spontaneously breaking the reflection symmetry of the lattice. Elementary excitations are spinons of two flavors, localized kinks and mobile antikinks. The gap to spin-1 excitations, $\Delta = 0.215J$ is determined by the edge of the two-spinon continuum. The introduction of a DM term with the $\mathbf D$ vector pointing along the same axis for all bonds, Fig.~\ref{fig:sawtooth}, lowers the spin-rotation symmetry down to an O(2). At weak coupling $D$, the lattice reflection symmetry remains spontaneously broken. At the same time, a finite $D$ lowers the excitation energies of both kinks and antikinks and the spin gap (understood as the lowest energy of $S_z=1$ excitations) begins to close.  A fairly crude analytical calculation indicates that the main factor affecting the spin gap is the minimum energy of the kink, $-2|D|$. The gap closes roughly when that energy equals the initial gap in absolute terms, $|D| = D_c \approx \Delta/2 \approx 0.1 J$.  This is confirmed by numerical work involving exact diagonalization of finite chains, with the result $D_c = 0.115J$.  Beyond the critical coupling, the spinons proliferate. Since they act as domain walls in the valence-bond order parameter, the valence-bond order is lost and the lattice symmetry is fully restored. The resulting state is likely a Luttinger liquid with incommensurate spin correlations and spin-wave excitations. Similar transitions between Ising-ordered phases and Luttinger liquids have been found in other one-dimensional systems.\cite{PhysRevB.18.421, fouet:2006prb} The strength of the DM coupling $D \approx (\delta g/g)J$ where $\delta g$ is the deviation of gyromagnetic ratio from its free-electron value $g$.\cite{PhysRev.120.91} In kagome antiferromagnets herbertsmithite and volborthite, $\delta g/g\approx 0.1$.\cite{jpcs.145.012002}

It is tempting to speculate that a somewhat similar transition may occur in the $S=1/2$ Heisenberg model on kagome with a DM coupling. While the existence of the transition is not in doubt---at a large enough $D$ the system should develop magnetic order\cite{cepas:140405, PhysRevB.79.214415, PhysRevB.81.064428, PhysRevB.81.144432}---the nature of the transition remains to be determined.

In the kagome antiferromagnet, spinon excitations are very similar to those of the sawtooth chain.\cite{PhysRevB.81.214445} In the absence of the DM term, kinks are localized and have zero energy, whereas antikinks follow one-dimensional trajectories with the same energetics as on the sawtooth chain. Adding the DM term thus has similar consequences, namely delocalization of kinks is the main factor lowering the edge of the kink-antikink continuum.  If anything, the gap may close even faster than on the sawtooth chain because on kagome kinks move in two dimensions and thus can lower their energy through delocalization more effectively than on a one-dimensional chain. For this reason, the critical DM coupling for kagome may be even lower than for the sawtooth chain.

The kagome antiferromagnet differs from the sawtooth chain in one important respect: it has a finite concentration of antikinks in the ground state. The antikinks form tightly bound $S=0$ pairs, whose binding energy $\Delta_\mathrm{aa} \approx 0.06J$ is lower than the threshold energy of kink-antikink creation $\Delta_\mathrm{ka} \approx 0.25J$.  Therefore the spin gap in the Heisenberg antiferromagnet on kagome is determined by binding energy of an antikink pair. Although the binding energy $\Delta_\mathrm{aa}$ is no doubt influenced by the introduction of the DM term, it is unlikely that this energy is very sensitive to the presence of a small perturbtion like $D$ as $\Delta_\mathrm{aa}$ is determined by a competition of two high-energy processes: the antikink hopping amplitude and the antikink attraction in the singlet channel, both with a strength of order $J$. It seems more likely that the larger gap $\Delta_{ka}$ will be quickly driven to zero as it is on the sawtooth c
 hain.

The nature of the phase transition at the conjectured condensation of kinks and antikinks is an open question. It is not even known whether the $D=0$ ground state is a valence-bond liquid or solid, with contradictory indications from different numerical techniques.\cite{singh:180407, PhysRevLett.104.187203, jiang:117203, Science.332.1173} (In our view, even a small amount of bond disorder will turn the system into a disordered valence-bond solid.) Adding the DM term will tend to melt the delicate valence-bond order turning the valence-bond crystal into a liquid before the magnetic condensation and thus inducing another phase transition along the way.  The nature of the condensed phase is not clear, either. Usually, ordering of the transverse components of magnetization is associated with a proliferation of $S_z=1$ objects, as is the case in magnon condensation,\cite{bec-review:2008} whereas here the condensing particles are spinons with half-integer spin.  This obsrvation
 lends support to the scenario with an intermediate gapless phase lacking long-range spin order,\cite{PhysRevB.79.214415} which is some sort of an algebraic spin liquid.\cite{ryu:184406}

\section*{Acknowledgments}

Work at JHU was supported in part by the U.S. Department of Energy, Office of Basic Energy Sciences, Division of Materials Sciences and Engineering under Award No. DE-FG02-08ER46544. OT and ZH acknowledge hospitality of the Max-Planck-Institute for Physics of Complex Systems, where they were part of the Advanced Study Group on Unconventional Magnetism in High Fields. YW was supported by William Gardner fellowship.
AS and JW were supported  by the National Science Foundation under NSF Grant No. DMR-0907793, and would like to thank I. McCulloch for patiently explaining the Matrix Product Toolkit.

\appendix
\section{Oscillations in the ground-state splitting}

To understand the oscillatory behavior of the ground-state splitting, Eq.~(\ref{eq:scaling}), we turn to a much simpler model: the antiferromagnetic XXZ chain with DM interaction described by the Hamiltonian $H = H_\mathrm{XXZ}+H_\mathrm{DM}$, where
\begin{equation}
    H_\mathrm{XXZ} = \sum_{n}\left[J\cos{\alpha}(S^x_n S^x_{n+1} + S^y_n S^y_{n+1})+J_zS^z_n S^z_{n+1}\right]
\end{equation}
and
\begin{equation}
    H_\mathrm{DM} = J\sin{\alpha}\sum_{n}(S^x_n S^y_{n+1} - S^y_n S^x_{n+1}).
\end{equation}
In the easy-axis limit, $J_z \gg J$, the ground state is doubly degenerate and exhibits N\'eel order. In a finite chain with periodic boundary condition, quantum tunneling splits the doublet into eigenstates with momenta 0 and $\pi$.  Below we discuss the effect of the DM term, $\alpha \neq 0$, on the splitting.

By rotating local axes at site $n$ through angle $n\alpha$ in the $xy$ plane, the DM term in the Hamiltonian can be removed, producing the standard XXZ model:
\begin{equation}
    H' = \sum_{n}\left[J(S^x_n S^x_{n+1} + S^y_n S^y_{n+1})+J_zS^z_n S^z_{n+1}\right].
\end{equation}
For a closed chain of length $L$, the transformation yields twisted periodic boundary conditions:
\begin{eqnarray}
    S^x_N &=& S^x_0\cos{(L\alpha)}+S^y_0\sin{(L\alpha)},
    \nonumber\\
    S^y_N &=& -S^x_0\sin{(L\alpha)}+S^y_0\cos{(L\alpha)}.
\end{eqnarray}
The twist is absent if $L\alpha = 2\pi m$, where $m$ is an integer.  Then the system has the same spectrum as in the absence of the DM term, $\alpha = 0$. At a fixed chain length $L$, the splitting is a periodic function of $\alpha$ with a period of $2\pi/L$.

To see that the splitting should have an oscillatory character, consider the special case of a $\pi$ twist, $L\alpha = (2m+1)\pi$. As Haldane argued,\cite{haldane:1983} the tunneling between the two N\'eel states is mediated by instantons with quantized winding numbers $n$, classical action $S_n$, and a Berry phase $\exp{(2\pi i nS)}$. For boundary conditions with a $\pi$ twist, the winding numbers are half-integer, $n = \pm 1/2, \pm 3/2, \ldots$ Instantons with opposite winding numbers have the same classical action, $S_n = S_{-n}$.  However, their Berry phases are exactly opposite, $\exp{(2\pi i nS)} = - \exp{(-2\pi i nS)}$, when both the winding numbers $n$ and spin $S$ are half-integer. As a result of destructive interference of instantons with opposite winding numbers, the tunneling amplitude vanishes when $L\alpha = (2m+1)\pi$.  We thus expect an oscillatory dependence of the splitting on $\alpha$ at a constant $L$ in the XXZ chain with half-integer spins and periodic b
 oundary conditions.  The exponential dependence of the splitting on the length will acquire an oscillating prefactor $\cos{(\alpha L)}$. This inspired Eq.~(\ref{eq:scaling}).

\section{Spin wave in sawtooth chain}
We compute the spin-wave spectrum on the sawtooth chain in the classical limit, $S \to \infty$. The Hamiltonian is
\begin{equation}\label{hami}
    H=\sum_{\langle ij \rangle}[\mathbf{S}_i \cdot \mathbf{S}_{j}
    	+ \mathbf D_{ij} \cdot(\mathbf{S}_i\times\mathbf{S}_j)].
\end{equation}
For brevity, we set $J=1$.

In equilibrium, spins lie in the plane normal to the DM vectors $\mathbf D_{ij}$, with the angle of $120^\circ$ between nearest neighbors, Fig.~\ref{fig:sawtooth}(f). It is convenient to choose reference frames in such a way that spins point along the local $z$ axes, the $x$ axes are in the plane of the spins, and the $y$ axes are parallel to $\mathbf D_{ij}$. For small deviations from equilibrium,
\begin{equation}\label{para}
\mathbf{S}_i
\approx S(\alpha_i, \, \beta_i, \, 1-\alpha_i^2/2-\beta_i^2/2)
\end{equation}
where $\alpha_i$ and $\beta_i$ are small deviations from the $120^\circ$ pattern.

In the harmonic approximation, the energy (\ref{hami}) reads
\begin{equation}\label{hami_simple}
    H = S^2\sum_{\langle ij \rangle}(\chi\beta_i\beta_j + \alpha_i\alpha_j)
    	-S^2\sum_{i}K_i \chi(\alpha_i^2+\beta_i^2)
\end{equation}
where $\chi=-1/2-\sqrt{3}D/2$. $K_i=1$ if $i$ is an apex (A) site and $K_i=2$ if it is a base (B) site.

The dynamics can be obtained from the Lagrangian, which includes a Berry phase term in addition to the potential energy:
\begin{equation}\label{lag}
    L=S\sum_i (\cos\theta_i-1)\dot{\phi}_i - H.
\end{equation}
After expressing the angles $\theta$ and $\phi$ in terms of $\alpha$ and $\beta$,
\begin{equation}\label{coord}
    \tan\phi = \beta/\alpha,
    \qquad
    \cos\theta \approx 1 - (\alpha^2+\beta^2)/2,
\end{equation}
we obtain the following Lagrangian:
\begin{equation}\label{lag_new}
    L = S\sum_i(\dot\alpha_i \beta_i - \alpha_i \dot \beta_i)/2 - H.
\end{equation}
It yields the equations of motion for spins on sublattices $A$ and $B$:
\begin{subequations}\label{equationofmotion}
\begin{eqnarray}
    \dot{\alpha}^A_{i} &=& S\chi(\beta^B_{i+1/2}+\beta^B_{i-1/2}) - 2S\chi\beta^A_i,\\
    \dot{\alpha}^B_i &=& S\chi(\beta^A_{i + 1/2}+\beta^A_{i-1/2}+\beta^B_{i+1}+\beta^B_{i-1}) - 4S\chi\beta^B_i,\\
    \dot{\beta}^A_i &=& -S(\alpha^B_{i+1/2}+\alpha^B_{i-1/2}) + 2S\chi\alpha^A_i,\\
    \dot{\beta}^B_i &=& -S(\alpha^A_{i+1/2}+\alpha^A_{i-1/2}+\alpha^B_{i+1}+\alpha^B_{i-1}) + 4S\chi\alpha^B_i.    
\end{eqnarray}
\end{subequations}
Note that $i$ is half-integer on sublattice A and integer on sublattice B. Plane waves with frequency $\omega$ and wavevector $k$ satisfy the equation
\begin{equation}\label{momentum}
    \left(\begin{array}{cccc}
    -i\omega & 0& 2S\chi &-2S\chi\cos(k/2)\\
    0&-i\omega&-2S\chi\cos(k/2) &4S \chi -2S \chi\cos{k}\\
    -2S\chi &2S\cos(k/2)&-i\omega &0\\
    2S\cos(k/2)& -4S\chi+2S\cos{k} &0&-i\omega\end{array}\right)\left(\begin{array}{c}\alpha^A\\
    \alpha^B\\ \beta^A\\ \beta^B
    \end{array}\right)=0.
\end{equation}
At $D=0$, we have one zero mode and one mode with a finite frequency $\omega = S\sqrt{2-\cos(2k)}$.  For a finite $D$, the zero mode acquires a dispersion linear in $k$ in the limit $k \to 0$. The wave velocity is
\begin{equation}\label{hd}
    v = 3S\sqrt{\frac{\sqrt{3}D+7D^2 +5\sqrt{3}D^3+3D^4}{2+8\sqrt{3}D+18D^2}}.
\end{equation}
Restoring $J$ as a coupling constant, we find the following behavior for the velocity.
As $D\to 0$, $v \sim 2.79S\sqrt{DJ}$. For $D=0.19J$, $v = 1.05SJ$.

\bibliography{kagome,quantum}
\end{document}